\pdfoutput=1
\documentclass[prl,aps,preprintnumbers,floats,floatfix,superscriptaddress,preprintnumbers,showpacs,eqsecnum,nofootinbib,notitlepage,twocolumn]{revtex4-2}
\usepackage[utf8]{inputenc}
\usepackage{latexsym,array,theorem,mathrsfs,bm,float}
\usepackage{psfrag}
\usepackage{amsfonts,amsmath,amssymb,afterpage,epsfig,color,graphicx,tabularx,here,multirow}

\newcommand{\calR}{{\cal R}}
\newcommand{\Mpl}{M_{\rm Pl}}

\newcommand{\ncpl}{{\alpha}}

\newcommand{\D}{{\rm d}}

\begin{document}
\baselineskip=12pt

%<<<<<<<<<<<<< TITLE >>>>>>>>>>>>>>>%
%%
\preprint{YITP-23-92} 
\title{Extra-tensor-induced origin for the PTA signal:\\
	No primordial black hole production}
%%
%<<<<<<<<<<<<< AUTHOR >>>>>>>>>>>>>>>%
%%
\author{Mohammad Ali Gorji}
\email{gorji@icc.ub.edu}
\affiliation{Departament de F\'{i}sica Qu\`{a}ntica i Astrof\'{i}sica, Institut de Ci\`{e}ncies del Cosmos, Universitat de Barcelona, Mart\'{i} i Franqu\`{e}s 1, 08028 Barcelona, Spain }

\author{Misao Sasaki}
\email{misao.sasaki@ipmu.jp}
\affiliation{Kavli Institute for the Physics and Mathematics of the Universe (WPI), The University of Tokyo, 277-8583, Chiba, Japan}
\affiliation{Center for Gravitational Physics and Quantum Information, Yukawa Institute for Theoretical Physics, Kyoto University, Kyoto 606-8502, Japan}
\affiliation{Leung Center for Cosmology and Particle Astrophysics,
National Taiwan University, Taipei 10617, Taiwan }

\author{Teruaki Suyama}
\email{suyama@phys.titech.ac.jp}
\affiliation{Department of Physics, Tokyo Institute of Technology, 2-12-1 Ookayama, Meguro-ku, Tokyo 152-8551, Japan}

%<<<<<<<<<<<<< DATE >>>>>>>>>>>>>>>%
\date{\today}

%======================================%
%<<<<<<<<<<<<< ABSTRACT >>>>>>>>>>>>>>>%
%======================================%
\begin{center}
\begin{abstract}
Recently, pulsar timing array (PTA) collaborations announced evidence for an isotropic stochastic gravitational wave (GW) background. The origin of the PTA signal can be astrophysical or cosmological. In the latter case, the so-called secondary scalar-induced GW scenario is one of the viable explanations, but it has a potentially serious issue of the overproduction of primordial black holes (PBHs) due to the enhanced curvature perturbation.
In this letter, we present a new interpretation of the PTA signal. Namely, it is originated from an extra spectator tensor field that exists on top of the metric tensor perturbation. As the energy density of the extra tensor field is always subdominant, it cannot lead to the formation of PBHs. Thus our primordial-tensor-induced scenario is free from the PBH overproduction issue.
\end{abstract}
\end{center}

\maketitle

{\bf Introduction.} Very recently, many PTA collaborations, NANOGrav \cite{NANOGrav:2023gor}, EPTA/InPTA \cite{Antoniadis:2023rey}, PPTA \cite{Reardon:2023gzh}, and CPTA \cite{Xu:2023wog}, have announced that they found evidence for an isotropic stochastic GW background. The PTA signal may be of astrophysical or cosmological origin. The natural astrophysical candidate for the signal is the supermassive black holes. However, there is a mild tension between the theory and observations \cite{NANOGrav:2023hvm,NANOGrav:2023gor,NANOGrav:2020spf,Antoniadis:2023zhi}. On the other hand, many cosmological models can fit the PTA data better \cite{Figueroa:2023zhu,NANOGrav:2023hvm,Antoniadis:2023zhi}. Although it is premature to conclude that the signal is cosmological, it is worth clarifying which type of cosmological models are more favored by the PTA data. This is important at least for an obvious reason that it enables us to constrain or exclude certain types of cosmological models. 

Among a number of cosmological origin scenarios that have been discussed \cite{Madge:2023cak,Kitajima:2023cek,Vagnozzi:2023lwo,Zu:2023olm,Guo:2023hyp,Gouttenoire:2023ftk,Abe:2023yrw,Servant:2023mwt,Li:2023tdx,Geller:2023shn,Gouttenoire:2023bqy,Salvio:2023ynn,Babichev:2023pbf,Zhang:2023nrs,Ahmadvand:2023lpp},
the so-called secondary scalar-induced GW scenario has attracted special attention \cite{Franciolini:2023pbf,Inomata:2023zup,Wang:2023ost,Cai:2023dls,Yi:2023mbm,Zhu:2023faa,Firouzjahi:2023lzg,You:2023rmn,HosseiniMansoori:2023mqh,Balaji:2023ehk,Cheung:2023ihl,Jin:2023wri}, which is based on an enhancement of the curvature perturbation on small scales during inflation \cite{Ananda:2006af,Baumann:2007zm,Alabidi:2012ex,Espinosa:2018eve,Kohri:2018awv}. However, the enhancement of the curvature perturbation compatible with the PTA signal could lead to the overproduction of PBHs \cite{Franciolini:2023pbf,Inomata:2023zup,Wang:2023ost,HosseiniMansoori:2023mqh,Balaji:2023ehk,Cheung:2023ihl}. 
In any case, this is a nontrivial and model-dependent issue that should be carefully examined case by case. In this letter, in contrast to the scalar-induced GW scenario, we show that the PTA signal can be well explained by the primordial-tensor-induced scenario, which is completely free from the issue of PBH overproduction.\\
%%%%%%%%%%%%%%%%%%%%%%%%%%%%%%%%%%%%%%%%%%%%%%%%%%%%%%%%%%%%%%%%%%%%%%%%%%%%%%%%%%%%%%%%%%%%%%%%%%%%%%%%%%%%%%%%%%%%%%%%%%%%%%%%%%%%%%%%%%%%%%%%%%%%%%%%%%%%%%%%%%%%%%%%%%%%

{\bf Cosmological GW background.}
The equations of motion for the metric tensor perturbation, that characterizes GWs, in a spatially flat Friedmann‐Lema\^itre‐Robertson-Walker background are given by
\begin{equation}\label{EoM-iGWs}
h''_{ij} + 2 \frac{a'}{a} h'_{ij} - \partial^2 h_{ij} = S_{ij} \,,
\end{equation} 
where a prime denotes derivative with respect to the conformal time $\tau$, 
$a$ is the scale factor, $\partial^2=\delta^{ij}\,\partial^2/\partial x^i\partial x^j$ and $S_{ij}$ is the source,
which may be schematically expressed in the following expansion form
\begin{equation}\label{source-expansion}
S_{ij} = {\cal O}(\epsilon_T) + {\cal O}(\epsilon_S^2) + {\cal O}(\epsilon_V^2) +  {\cal O}(\epsilon_T^2) 
+ \cdots
\,,
\end{equation} 
where $\epsilon_S$, $\epsilon_V$, and $\epsilon_T$ represent the amplitudes of scalar, vector, and extra tensor perturbations. An extra tensor perturbation may exist on top of the metric tensor perturbation.

The spectral density fraction of GWs during the radiation dominance is given by
\begin{align}\label{GWs-sd}
&\Omega_{\rm GW}(k,\tau) = \frac{1}{12} \left(\frac{k}{aH}\right)^2
{{\cal P}_{h}(k,\tau)}
\,,
\end{align}
where the power spectrum of the metric tensor perturbation is defined as $\langle {h}^\lambda_{\bf k}(\tau) {h}^{\ast{r}}_{{\bf q}}(\tau) \rangle 
= ({2\pi^2}/{k^3}) {\cal P}^\lambda_h(k,\tau) \delta^{\lambda{r}} \delta({\bf k}+{\bf q})$ with ${\cal P}_{h}(k,\tau) = \sum_{\lambda}{\cal P}^\lambda_{h}(k,\tau)$. Here ${h}^\lambda_{\bf k}$ is the Fourier amplitude of $h_{ij}$ that is defined as usual $h_{ij}(\tau,{\bf k}) = \sum_{\lambda=+,\times} e^{\lambda}_{ij}(\hat{\bf k}) h^{\lambda}_{\bf k}(\tau)$, where $e_{ij}^\lambda(\hat{\bf k})$ is the linear polarization tensor \cite{Caprini:2018mtu}.
Ignoring a time variation of the relativistic degrees of freedom,
$\Omega_{\rm GW}(k,\tau)$ is independent of time during the radiation dominated era.

The energy density of GWs today is related to that of GWs during the radiation dominance, $\Omega_{{\rm GW},{\rm r}}(k)$, as
\begin{align}
\Omega_{{\rm GW,0}}(k) h^2
&\sim  \Omega_{{\rm r},0} h^2 \Omega_{{\rm GW},{\rm r}}(k)
\nonumber \\ \label{Omega-GWs}
&\sim 10^{-5} \Omega_{{\rm GW},{\rm r}}(k) \,,
\end{align}
where $\Omega_{{\rm r},0} $ is the fraction of energy density of radiation today
and $h=H_0/(100~{\rm km/Mpc/s})$. 
%The subscripts ``$0$" and ``${\rm em}$" denote today and the time of emission of GWs respectively. In obtaining the numerical prefactor in the second line, we have used the values $g_{\rm s,0}=3.91$, $g_{\rho,0}=2$, $g_{{\rm s, em}}\sim g_{\rho,{\rm em}}=106.75$ and $\Omega_{{\rm r},0} h^2=4.2\times10^{-5}$.

It is useful to first look at the contribution of the vacuum fluctuations of the metric tensor perturbation to the stochastic GW background \eqref{Omega-GWs}, which corresponds to the case when the source term in Eq. \eqref{EoM-iGWs} vanishes. On CMB scales $k_{\rm CMB} \sim 0.05\, \mbox{Mpc}^{-1}$, the power spectrum of the metric tensor perturbation is constrained as $r_{\rm CMB}={\cal P}^{\rm CMB}_h/{\cal P}^{\rm CMB}_\calR<10^{-2}$ where $r_{\rm CMB}$
is the tensor-to-scalar ratio and ${\cal P}^{\rm CMB}_\calR$ is the power spectrum of the curvature perturbation $\calR$.\footnote{More precisely, the CMB data put bound $r_{\rm CMB}<0.056$ \cite{Planck:2018jri} or $r_{\rm CMB}<0.034$ \cite{Tristram:2021tvh} on the tensor-to-scalar ratio. Here, for our rough estimation, we consider $r_{\rm CMB}<10^{-2}$.} Taking into account the CMB constraint ${\cal P}^{\rm CMB}_{\calR}\sim10^{-9}$ \cite{Planck:2018jri}, we find ${\cal P}^{\rm CMB}_h<10^{-11}$. Assuming almost scale-invariant ${\cal P}_h$, the amplitude does not change significantly at smaller scales and therefore ${\cal P}^{\rm PTA}_h<10^{-12}$, where $k_{\rm PTA}\sim10^6\, \mbox{Mpc}^{-1}$ corresponds to the PTA scale/frequency. 
%Taking into account that ${\cal P}_{h,{\rm r}}(k,\tau)\sim \left(a_{\rm r}/a_{\ast}\right)^2{\cal P}_{h}(k)$ 
Evaluating the r.h.s. of Eq.~(\ref{GWs-sd}) at the time of horizon-reentry $k=aH$ for which we can approximately replace ${\cal P}_h(k,\tau)$ 
with the initial power spectrum ${\cal P}_h(k)$ 
given on super-horizon scale, we find $\Omega_{{\rm GW},{\rm r}}(k)\sim{\cal P}_{h}(k)$.
%, where ${\cal P}_{h,{\rm r}}(k,\tau)$ and ${\cal P}_{h}(k)$ 
%are the power spectra deep inside the horizon and at superhorizon during radiation dominance respectively and $a_{\rm r}$ and $a_{\ast}$ are the corresponding scale factors, from Eq. \eqref{GWs-sd} we find $\Omega_{{\rm GW},{\rm r}}(k)\sim{\cal P}_{h}(k)$. 
Then, Eq. \eqref{Omega-GWs} gives $\Omega^{\rm PTA}_{{\rm GW,0}}(k) h^2\sim10^{-5}{\cal P}^{\rm PTA}_h<10^{-17}$.  However, the amplitude of the PTA signal is much stronger \cite{NANOGrav:2023hvm,Antoniadis:2023zhi,Figueroa:2023zhu}\footnote{The signal has larger amplitude for large frequencies and smaller amplitude at smaller frequencies. However, for our purpose of estimation, we consider an average value which does not change significantly our final conclusion.}
\begin{align}\label{PTA-signal}
10^{-9} \lesssim \Omega^{\rm PTA}_{{\rm GW,0}} h^2\lesssim 10^{-7}\,.
\end{align}
Thus, the amplitude of GWs from the vacuum metric tensor fluctuations during inflation is much smaller than the one observed by PTA. We, therefore, need to look for an alternative scenario that provides an appropriate source term for Eq. \eqref{EoM-iGWs}. \\

%%%%%%%%%%%%%%%%%%%%%%%%%%%%%%%%%%%%%%%%%%%%%%%%%%%%%%%%%%%%%%%%%%%%%%%%%%%%%%%%%%%%%%%%%%%%%%%%%%%%%%%%%%%%%%%%%%%%%%%%%%%%%%%%%%%%%%%%%%%%%%%%%%%%%%%%%%%%%%%%%%%%%%%%%%%%

{\bf Secondary scalar-induced GWs.} In the absence of any extra tensor perturbations, the curvature perturbation $\calR$,
corresponding to ${\cal O}(\epsilon_S^2)$ in Eq. \eqref{source-expansion},
gives the dominant contribution to the r.h.s of Eq. \eqref{EoM-iGWs} \cite{Ananda:2006af,Baumann:2007zm,Saito:2008jc,Alabidi:2012ex,Espinosa:2018eve,Kohri:2018awv,Pi:2020otn} leading to the usual scenario of the secondary scalar-induced GWs (see \cite{Domenech:2021ztg} for a review). Looking at Eq. \eqref{EoM-iGWs}, we see that two-point function of $h_{ij}$ is proportional to the four-point function of $\calR$. Assuming Gaussianity of $\calR$, we then find
\begin{align}
{\cal P}^{\rm S.I.}_{h}(k)\sim{\cal P}^2_{\calR}(k) \,,
\end{align}
which after substituting in \eqref{Omega-GWs} gives the following result for the scalar-induced GW spectrum today,
\begin{align}\label{GWs-SI}
\Omega^{\rm S.I.}_{{\rm GW,0}}(k) h^2
\sim 10^{-5} {\cal P}_{\calR}^2 \,.
\end{align}
On CMB scales $k_{\rm CMB} \sim 0.05\, \mbox{Mpc}^{-1}$, ${\cal P}^{\rm CMB}_{\calR}\sim 10^{-9}$ \cite{Planck:2018jri}.
If we simply extrapolate this magnitude of the powerspectrum down to the PTA scale $k_{\rm PTA}\sim10^6\, \mbox{Mpc}^{-1}$, we find $\Omega^{\rm S.I.}_{{\rm GW,0}}(k_{\rm PTA}) h^2\sim 10^{-23}$ which is too small to explain the PTA signal. Indeed, it is too small to be detected even with the future GW detectors. However, the PTA signal is observed at much smaller scale %$k_{\rm PTA}\sim10^6\, \mbox{Mpc}^{-1}$. 
and ${\cal P}_\calR$ at the PTA scale may significantly deviate from that at CMB scales. Assuming that power spectrum is enhanced on small scales as $10^{-1}\lesssim{\cal P}^{\rm PTA}_{\calR}\lesssim10^{-2}$, one can explain the PTA signal in scalar-induced GW scenarios \cite{Franciolini:2023pbf,Inomata:2023zup,Wang:2023ost,Cai:2023dls,Yi:2023mbm,Zhu:2023faa,Firouzjahi:2023lzg,You:2023rmn,HosseiniMansoori:2023mqh}.

Assuming Gaussian distribution for $\calR$, the large values $10^{-1}\lesssim{\cal P}^{\rm PTA}_{\calR}\lesssim10^{-2}$ may lead to the overproduction of PBHs and scalar-induced scenario may fail to explain the PTA signal \cite{Franciolini:2023pbf}. Although this is a model-dependent issue \cite{Sasaki:2018dmp,Yoo:2018kvb,Germani:2018jgr,Germani:2019zez} and the role of non-Gaussianity and other parameters in the model can be very important \cite{Saito:2008em,Young:2013oia,Franciolini:2018vbk,Atal:2018neu,Cai:2018dig,Passaglia:2018ixg,DeLuca:2019qsy,Atal:2019cdz,Suyama:2019cst,Yoo:2019pma,Atal:2019erb,Ezquiaga:2019ftu,Cai:2021zsp,Pi:2022ysn,Franciolini:2023pbf,Inomata:2023zup,Wang:2023ost,Liu:2023ymk,Balaji:2023ehk}, a careful PBH analysis is necessary whenever the PTA signal is interpreted by the scalar-induced GW scenario.\\

{\bf Primordial-tensor-induced GWs.} Now, let us review the primordial-tensor-induced scenario which is the main focus of this letter. The contribution of a spectator field to the GWs is usually assumed to be negligible compared with the one coming from the curvature perturbation that corresponds to ${\cal O}(\epsilon_S^2)$ in the source \eqref{source-expansion}. As it is pointed out in~\cite{Gorji:2023ziy}, this is not the case when there is a spectator field that provides an extra tensor perturbation, as it may show up at its linear order ${\cal O}(\epsilon_T)$ in the source \eqref{source-expansion}. Therefore, even if the amplitude of the spectator tensor perturbation is smaller than that of the curvature perturbation ${\cal O}(\epsilon_T)\ll{\cal O}(\epsilon_S)$, it will be dominant  when ${\cal O}(\epsilon_T)\gg{\cal O}(\epsilon_S^2)$. This is the key feature of the primordial-tensor-induced GW scenario \cite{Gorji:2023ziy}. 
%Looking at Eq. \eqref{EoM-iGWs} \su{[Teruaki: While Eq.(8) already assumes the conversion efficiency from $t_{ij}$ to $h_{ij}$ is ${\cal O}(1)$, Eq.(9) does not. So I found it confusing to present Eq.(8) and related argument. Maybe, we can eliminate the part "Looking at Eq.(1), ..... and we will clarify this point later in this section."]}, we see that, in tensor-induced scenario, the two-point function of $h_{ij}$ is proportional to the two-point function of the extra tensor perturbations $t_{ij}$ and we find
%\begin{align}\label{Ph-TI}
%{\cal P}^{\rm T.I.}_{h}(k)\sim{\cal P}_{t}(k) \,,
%\end{align}
%where ${\cal P}_{t}(k)$ is the dimensionless power spectrum of $t_{ij}$ at superhorizon scales. Of course, the order of magnitude of the coefficient in Eq. \eqref{Ph-TI} depends on the model. Here, we have assumed that it is of the order of unity and we will clarify this point later in this section. 

In \cite{Gorji:2023ziy}, the effective field theory approach \cite{Bordin:2018pca} is implemented to study a linear system of massless extra tensor perturbation $t_{ij}$ that is coupled to $h_{ij}$. The quadratic action of the model is given by 
\begin{align}
	S &= \frac{\Mpl^2}{8} \int \D^3x\, \D\tau\, a^2 \left[ 
	\left( {h}'_{ij} \right)^2 
	- \left( \partial_i{h}_{jk} \right)^2
	\right]
	\nonumber \\ \nonumber
	&+
	\frac{1}{2} \int \D^3x\, \D\tau\, a^2 f^2 \left[ 
	\left( {t}'_{ij} \right)^2 
	- c_t^2 \left( \partial_i{t}_{jk} \right)^2
	\right]
	\\ \label{action}
	&+ \frac{\Mpl}{2}
	\int \D^3x\, \D\tau\, \ncpl\, a a' \left[ h'_{ij} t^{ij} \right] \,,
\end{align}
where $\Mpl=1/\sqrt{8\pi{G}}$ in unit $\hbar=1=c$, $c_t$ is the sound speed of $t_{ij}$, and $f$, $\ncpl$ are functions of time. The equations of motion in Fourier space are 
\begin{align}\nonumber
&h''^\lambda_{\bf k} + 2 \frac{a'}{a} h'^\lambda_{\bf k} + k^2 {h}^\lambda_{\bf k}
= - \frac{2\ncpl}{\Mpl} \frac{a'}{a}
\left[
t'^{\lambda}_{\bf k} + \frac{(\ncpl a a')'}{\ncpl a a'} {t}^{\lambda}_{\bf k} 
\right] \,,
\\ \nonumber
&t''^\lambda_{\bf k} + 2 \frac{(af)'}{af} t'^\lambda_{\bf k} + c_t^2k^2 {t}^\lambda_{\bf k}
= 
\frac{\Mpl\ncpl}{2f^2} \frac{a'}{a} h'^{\lambda}_{\bf k} \,.
\end{align}
Comparing the first equation above with Eq. \eqref{EoM-iGWs}, we clearly see that $t_{ij}$ provides a linear tensorial source for $h_{ij}$. 

To make the setup simple, we assume that $\ncpl$ vanishes during inflation $\ncpl_{\rm inf}=0$ while it is non-vanishing and constant during the radiation dominance $\ncpl_{\rm r}\neq0$. Thus, tensor-induced GWs only generate during radiation dominance. Note that one can easily extend our analysis to the case of
$\ncpl_{\rm inf}\neq0$. However, as it would only obscure the picture
by bringing in inessential technical complications, we focus on the
simple case of $\ncpl_{\rm inf}=0$. The time dependency of $c_t$ and $f$ makes it possible to enhance the amplitude of $t_{ij}$ on small sub-CMB scales $k={\cal O}(10^6-10^{18})\mbox{Mpc}^{-1}$ during inflation. Then, on superhorizon scales, ignoring the small contribution from the vacuum fluctuations of $h_{ij}$ in comparison with the enhanced amplitude of $t_{ij}$, the tensor-induced GW spectrum today was found as \cite{Gorji:2023ziy}
\begin{align}
\Omega^{\rm T.I.}_{{\rm GW,0}}(k) h^2
&\sim 10^{-5} {K_h(k,\tau)} \, 
\ncpl_{\rm r}^2 c_t^2 \frac{{\cal P}_{t}(k)}{3\Mpl^2} 
\nonumber \\ \label{GWs-TI}
&\sim 10^{-5} {K_h(k,\tau)} \, 
\ncpl_{\rm r}^2 \, \Omega_{t,{\rm r}}(k) \,,
\end{align}
where
\begin{align}
\Omega_{t,{\rm r}}(k) 
\sim \frac{c_t^2 {\cal P}_{t}(k)}{3\Mpl^2} \,,
\end{align}
is the fractional energy density of $t_{ij}$ during the radiation dominance. In the above expression, ${\cal P}_{t}(k)$ is the enhanced dimensionless power spectrum of $t_{ij}$ at the time of horizon re-entry. The kernel function $K_h$ takes into account the evolution of the enhanced metric tensor perturbation from the time of horizon re-entry to any time when its scale is deep inside the horizon. 
The explicit form of $K_h$ is given in \cite{Gorji:2023ziy}. 
Depending on the values of $c_t$ and $\ncpl_{\rm r}$, $K_h$ simplifies as
\begin{align}\label{K-h}
K_h \ncpl_{\rm r}^2 \sim \begin{cases}
\ncpl_{\rm r}^2(1-c_t^2){}^{-2} & c_t < 1 \,,
\\
\sin^2(\ncpl_{\rm r}\Delta{\cal N}/2) &  c_t = 1 \,.
\end{cases}
\end{align} 
Note that by $c_t<1$ we mean that $c_t$ is not very close to unity since in that case we need to use the result for $c_t=1$. The threshold can be found by looking at the value of $c_t$ for which the expression for $c_t<1$ and those for $c_t=1$ in Eq.~\eqref{K-h} become of the same order.

From Eq.~\eqref{K-h} it is clear that $K_h \ncpl_{\rm r}^2\leq1$ as far as $\ncpl_{\rm r}\leq1$. It is interesting to note that $K_h \ncpl_{\rm r}^2={\cal O}(1)$ can be achieved in the case of $c_t=1$ even for $\ncpl_{\rm r}\ll1$ which is needed from effective field theory point of view. This is possible if $\Delta{\cal N}$ is large enough such that $\ncpl_{\rm r}\Delta{\cal N}\sim1$ for $\ncpl_{\rm r}\ll1$. For instance, for the PTA frequency $k\sim10^6\,\mbox{Mpc}^{-1}$, $\Delta{\cal N}\sim20$ and $\Delta{\cal N}^{-1}\sim0.04$. Thus, $\ncpl_{\rm r}\Delta{\cal N}\sim1$ is possible for $\ncpl_{\rm r}\sim0.04\ll1$. More interestingly, $\ncpl_{\rm r}\Delta{\cal N}\gtrsim1$ is also possible for $0.04\lesssim\ncpl_{\rm r}<1$. For the latter case, as it can be seen from Eq. \eqref{K-h}, multiple peaks structure in the spectrum shows up, leading to an interesting phenomenology \cite{Gorji:2023ziy}. For our purpose of estimation of the order of magnitude, we assume $c_t\sim1$ and $K_h \ncpl_{\rm r}^2\sim1$ from now on.

Since we have assumed $\ncpl_{\rm inf}=0$, there is no linear coupling between $t_{ij}$ and $h_{ij}$ during inflation. Otherwise, the CMB constraints on the tensor-to-scalar ratio would give an upper bound on ${\cal P}^{\rm CMB}_{t}$ \cite{Maleknejad:2011jw,Dimastrogiovanni:2016fuu,Bordin:2018pca}. The only constraint that we have on CMB scales is that the energy density of $t_{ij}$ should be subdominant during inflation $\Omega_{t,{\rm inf}}\ll1$ or $c_t^2{\cal P}_t\ll M_{\rm Pl}^2$. This constraint can be easily satisfied for all wavenumbers at any time.

For the small scale modes $k={\cal O}(10^6-10^{18})\mbox{Mpc}^{-1}$, the extra tensor modes $t_{ij}$ contribute to the effective number of relativistic degrees of freedom, and $\Delta{N}_{\rm eff}<0.3$ \cite{Planck:2018vyg} gives the following big bang nucleosynthesis bound \cite{Gorji:2023ziy},
\begin{align}\label{BBN-bound}
&\Omega_{t,{\rm r}}(k) < 0.034\,;
&k\gtrsim{k}_{\rm BBN} \sim 10^4\,\mbox{Mpc}^{-1}\,.
\end{align}

Finally, there is an important theoretical bound of avoiding overproduction of PBHs which is the subject of the next section.\\

%%%%%%%%%%%%%%%%%%%%%%%%%%%%%%%%%%%%%%%%%%%%%%%%%%%%%%%%%%%%%%%%%%%%%%%%%%%%%%%%%%%%%%%%%%%%%%%%%%%%%%%%%%%%%%%%%%%%%%%%%%%%%%%%%%%%%%%%%%%%%%%%%%%%%%%%%%%%%%%%%%%%%%%%%%%%

{\bf PTA signal and PBHs formation.} Now we interpret the PTA signal to be originated from $t_{ij}$ in the tensor-induced scenario. Assuming that the enhancement in ${\cal P}_{t}(k)$ happens at small scales, i. e. $k_{\rm PTA}\sim 10^6\,\mbox{Mpc}^{-1}$,\footnote{Note that our model can completely match the frequency dependence of the PTA signal: we can achieve any shape for $\Omega_{t,{\rm r}}(k)$ or equivalently ${\cal P}_{t}(k)$ by an appropriate choice of the effective field theory free function $f$. } the question is whether the energy density of $t_{ij}$ around the time of horizon re-entry is large enough to form PBHs or not. This is characterized by the root-mean-square amplitude $\sigma^{\rm T.I.}_{\delta_{\rm r}}$ of the density contrast $\delta_{\rm r}$ due to $t_{ij}$, which is quadratic in $t_{ij}$.
% as shown in Eq. \eqref{energy-density-t-F1}. 
Thus we expect
\begin{align}\label{sigma-TI}
\sigma^{\rm T.I.}_{\delta_{\rm r}}
\ll \Omega_{t,{\rm r}} \,.
\end{align}
Applying the big bang nucleosynthesis constraint \eqref{BBN-bound} to
the above, we find
\begin{align}\label{PBHs}
\sigma^{\rm T.I.}_{\delta_{\rm r}} \ll 0.03 \,.
\end{align}
In order to not overproduce PBHs, we need to require $\sigma^{\rm
T.I.}_{\delta_{\rm r}}
\lesssim 0.04$ \cite{Harada:2013epa}. We see that this constraint is
safely satisfied.

Now, let us look for the value of $\Omega_{t,{\rm r}}$ which is
compatible with the PTA signal \eqref{PTA-signal}.
From Eqs. \eqref{PTA-signal} and \eqref{GWs-TI}, we find that the
tensor-induced GW interpretation of the PTA signal requires
\begin{align}\label{PTA}
& 10^{-4} \lesssim \Omega^{\rm PTA}_{t,{\rm r}} \lesssim 10^{-2} \,.
\end{align}
From \eqref{sigma-TI}, we see that the PTA signal implies $\sigma^{\rm
T.I.}_{\delta_{\rm r}}\ll 10^{-2}$.
Thus, the PTA signal can be explained in the tensor-induced GWs
scenario with totally negligible PBH production.
This result is easy to understand.
The extra tensor field whose energy density given by Eq. \eqref{PTA}
always remains as a spectator field. As such, it is almost impossible
for the density contrast to be large enough to form a black hole. This
is the main point of this letter.\\

%%%%%%%%%%%%%%%%%%%%%%%%%%%%%%%%%%%%%%%%%%%%%%%%%%%%%%%%%%%%%%%%%%%%%%%%%%%%%%%%%%%%%%%%%%%%%%%%%%%%%%%%%%%%%%%%%%%%%%%%%%%%%%%%%%%%%%%%%%%%%%%%%%%%%%%%%%%%%%%%%%%%%%%%%%%%

{\bf Summary.} Recently, many PTA collaborations have announced that they found evidence of an isotropic stochastic GW background. The origin of the signal can be either astronomical or cosmological. 
While it seems there is no convincing model of astronomical origin that can explain the data, many cosmological models have been discussed and examined under the PTA data. 
One of the models that fit the data well is the scalar-induced GW scenario, in which an enhancement of the curvature perturbation on small sub-CMB scales induces large amplitude GWs at second order in perturbation. However, in this scenario, the enhanced curvature perturbation potentially leads to an abundant PBH production, and it is suggested that the scenario may suffer from the PBH overproduction.
 
 In this letter, we have presented a new, viable interpretation in the context of the primordial-tensor-induced GW scenario. Namely, the PTA signal may have originated from an extra, spectator tensor field whose fluctuation amplitude has been enhanced during inflation. 
 Very interestingly, in addition to the fact that it passes major cosmological tests like BBN and CMB constraints, it is also completely free from the PBH overproduction.
 The reason is simple: the energy density of this extra tensor field remains always small in comparison with the dominant component of the universe at all times, and hence it can never give rise to a density contrast large enough to form PBHs. This is a distinct feature of the tensor-induced scenario, which makes it possible to distinguish it from the usual scalar-induced scenario.\\

%%%%%%%%%%%%%%%%%%%%%%%%%%%%%%%%%%%%%%%%%%%%%%%%%%%%%%%%%%%%%%%%%%%%%%%%%%%%%%%%%%%%%%%%%%%%%%%%%%%%%%%%%%%%%%%%%%%%%%%%%%%%%%%%%%%%%%%%%%%%%%%%%%%%%%%%%%%%%%%%%%%%%%%%%%%%

%%%%%%%%%%%%%%%%%%%%%%%%%%%%%%%%%%%%%%%%%%%%%%%%%%%%%%%%%%%%%%%
%%%%%%%%%%%%%%%%%%%%%%%%%%%%%%%%%%%%%%%%%%%%%%%%%%%%%%%%%%%%%%%
{\bf Acknowledgments.}
We thank Guillem Dom\`enech for an important comment on the draft and Shi Pi for useful discussions. MAG thanks Tokyo Institute of Techonology and Yukawa Institute for Theoretical Physics for hospitality and support when this work was in progress. MAG also thanks organizers of the workshop “IBS CTPU-CGA 2023 Summer Workshop and School on Particle Physics” in Korea where this work was in its final stage. The work of MAG was supported by Mar\'{i}a Zambrano fellowship.
This work was supported by the JSPS KAKENHI Grant Numbers JP19K03864 (TS), JP23K03411 (TS) and 20H05853 (MS), and by the World Premier International Research Center Initiative (WPI Initiative), MEXT, Japan. 
%\end{acknowledgements}
\\

%%%%%%%%%%%%%%%%%%%%%%%%%%%%%%%%%%%%%%%%%%%%%%%%%%%%%%%%%%%%%%%
%%%%%%%%%%%%%%%%%%%%%%%%%%%%%%%%%%%%%%%%%%%%%%%%%%%%%%%%%%%%%%%

\bibliography{ref}

\begin{thebibliography}{71}%
\makeatletter
\providecommand \@ifxundefined [1]{%
 \@ifx{#1\undefined}
}%
\providecommand \@ifnum [1]{%
 \ifnum #1\expandafter \@firstoftwo
 \else \expandafter \@secondoftwo
 \fi
}%
\providecommand \@ifx [1]{%
 \ifx #1\expandafter \@firstoftwo
 \else \expandafter \@secondoftwo
 \fi
}%
\providecommand \natexlab [1]{#1}%
\providecommand \enquote  [1]{``#1''}%
\providecommand \bibnamefont  [1]{#1}%
\providecommand \bibfnamefont [1]{#1}%
\providecommand \citenamefont [1]{#1}%
\providecommand \href@noop [0]{\@secondoftwo}%
\providecommand \href [0]{\begingroup \@sanitize@url \@href}%
\providecommand \@href[1]{\@@startlink{#1}\@@href}%
\providecommand \@@href[1]{\endgroup#1\@@endlink}%
\providecommand \@sanitize@url [0]{\catcode `\\12\catcode `\$12\catcode
  `\&12\catcode `\#12\catcode `\^12\catcode `\_12\catcode `\%12\relax}%
\providecommand \@@startlink[1]{}%
\providecommand \@@endlink[0]{}%
\providecommand \url  [0]{\begingroup\@sanitize@url \@url }%
\providecommand \@url [1]{\endgroup\@href {#1}{\urlprefix }}%
\providecommand \urlprefix  [0]{URL }%
\providecommand \Eprint [0]{\href }%
\providecommand \doibase [0]{http://dx.doi.org/}%
\providecommand \selectlanguage [0]{\@gobble}%
\providecommand \bibinfo  [0]{\@secondoftwo}%
\providecommand \bibfield  [0]{\@secondoftwo}%
\providecommand \translation [1]{[#1]}%
\providecommand \BibitemOpen [0]{}%
\providecommand \bibitemStop [0]{}%
\providecommand \bibitemNoStop [0]{.\EOS\space}%
\providecommand \EOS [0]{\spacefactor3000\relax}%
\providecommand \BibitemShut  [1]{\csname bibitem#1\endcsname}%
\let\auto@bib@innerbib\@empty
%</preamble>
\bibitem [{\citenamefont {Agazie}\ \emph {et~al.}(2023)\citenamefont {Agazie}
  \emph {et~al.}}]{NANOGrav:2023gor}%
  \BibitemOpen
  \bibfield  {author} {\bibinfo {author} {\bibfnamefont {G.}~\bibnamefont
  {Agazie}} \emph {et~al.} (\bibinfo {collaboration} {NANOGrav}),\ }\href
  {\doibase 10.3847/2041-8213/acdac6} {\bibfield  {journal} {\bibinfo
  {journal} {Astrophys. J. Lett.}\ }\textbf {\bibinfo {volume} {951}},\
  \bibinfo {pages} {L8} (\bibinfo {year} {2023})},\ \Eprint
  {http://arxiv.org/abs/2306.16213} {arXiv:2306.16213 [astro-ph.HE]}
  \BibitemShut {NoStop}%
\bibitem [{\citenamefont {Antoniadis}\ \emph
  {et~al.}(2023{\natexlab{a}})\citenamefont {Antoniadis} \emph
  {et~al.}}]{Antoniadis:2023rey}%
  \BibitemOpen
  \bibfield  {author} {\bibinfo {author} {\bibfnamefont {J.}~\bibnamefont
  {Antoniadis}} \emph {et~al.},\ }\href@noop {} {\  (\bibinfo {year}
  {2023}{\natexlab{a}})},\ \Eprint {http://arxiv.org/abs/2306.16214}
  {arXiv:2306.16214 [astro-ph.HE]} \BibitemShut {NoStop}%
\bibitem [{\citenamefont {Reardon}\ \emph {et~al.}(2023)\citenamefont {Reardon}
  \emph {et~al.}}]{Reardon:2023gzh}%
  \BibitemOpen
  \bibfield  {author} {\bibinfo {author} {\bibfnamefont {D.~J.}\ \bibnamefont
  {Reardon}} \emph {et~al.},\ }\href {\doibase 10.3847/2041-8213/acdd02}
  {\bibfield  {journal} {\bibinfo  {journal} {Astrophys. J. Lett.}\ }\textbf
  {\bibinfo {volume} {951}},\ \bibinfo {pages} {L6} (\bibinfo {year} {2023})},\
  \Eprint {http://arxiv.org/abs/2306.16215} {arXiv:2306.16215 [astro-ph.HE]}
  \BibitemShut {NoStop}%
\bibitem [{\citenamefont {Xu}\ \emph {et~al.}(2023)\citenamefont {Xu} \emph
  {et~al.}}]{Xu:2023wog}%
  \BibitemOpen
  \bibfield  {author} {\bibinfo {author} {\bibfnamefont {H.}~\bibnamefont {Xu}}
  \emph {et~al.},\ }\href {\doibase 10.1088/1674-4527/acdfa5} {\bibfield
  {journal} {\bibinfo  {journal} {Res. Astron. Astrophys.}\ }\textbf {\bibinfo
  {volume} {23}},\ \bibinfo {pages} {075024} (\bibinfo {year} {2023})},\
  \Eprint {http://arxiv.org/abs/2306.16216} {arXiv:2306.16216 [astro-ph.HE]}
  \BibitemShut {NoStop}%
\bibitem [{\citenamefont {Afzal}\ \emph {et~al.}(2023)\citenamefont {Afzal}
  \emph {et~al.}}]{NANOGrav:2023hvm}%
  \BibitemOpen
  \bibfield  {author} {\bibinfo {author} {\bibfnamefont {A.}~\bibnamefont
  {Afzal}} \emph {et~al.} (\bibinfo {collaboration} {NANOGrav}),\ }\href
  {\doibase 10.3847/2041-8213/acdc91} {\bibfield  {journal} {\bibinfo
  {journal} {Astrophys. J. Lett.}\ }\textbf {\bibinfo {volume} {951}},\
  \bibinfo {pages} {L11} (\bibinfo {year} {2023})},\ \Eprint
  {http://arxiv.org/abs/2306.16219} {arXiv:2306.16219 [astro-ph.HE]}
  \BibitemShut {NoStop}%
\bibitem [{\citenamefont {Pol}\ \emph {et~al.}(2021)\citenamefont {Pol} \emph
  {et~al.}}]{NANOGrav:2020spf}%
  \BibitemOpen
  \bibfield  {author} {\bibinfo {author} {\bibfnamefont {N.~S.}\ \bibnamefont
  {Pol}} \emph {et~al.} (\bibinfo {collaboration} {NANOGrav}),\ }\href
  {\doibase 10.3847/2041-8213/abf2c9} {\bibfield  {journal} {\bibinfo
  {journal} {Astrophys. J. Lett.}\ }\textbf {\bibinfo {volume} {911}},\
  \bibinfo {pages} {L34} (\bibinfo {year} {2021})},\ \Eprint
  {http://arxiv.org/abs/2010.11950} {arXiv:2010.11950 [astro-ph.HE]}
  \BibitemShut {NoStop}%
\bibitem [{\citenamefont {Antoniadis}\ \emph
  {et~al.}(2023{\natexlab{b}})\citenamefont {Antoniadis} \emph
  {et~al.}}]{Antoniadis:2023zhi}%
  \BibitemOpen
  \bibfield  {author} {\bibinfo {author} {\bibfnamefont {J.}~\bibnamefont
  {Antoniadis}} \emph {et~al.},\ }\href@noop {} {\  (\bibinfo {year}
  {2023}{\natexlab{b}})},\ \Eprint {http://arxiv.org/abs/2306.16227}
  {arXiv:2306.16227 [astro-ph.CO]} \BibitemShut {NoStop}%
\bibitem [{\citenamefont {Figueroa}\ \emph {et~al.}(2023)\citenamefont
  {Figueroa}, \citenamefont {Pieroni}, \citenamefont {Ricciardone},\ and\
  \citenamefont {Simakachorn}}]{Figueroa:2023zhu}%
  \BibitemOpen
  \bibfield  {author} {\bibinfo {author} {\bibfnamefont {D.~G.}\ \bibnamefont
  {Figueroa}}, \bibinfo {author} {\bibfnamefont {M.}~\bibnamefont {Pieroni}},
  \bibinfo {author} {\bibfnamefont {A.}~\bibnamefont {Ricciardone}}, \ and\
  \bibinfo {author} {\bibfnamefont {P.}~\bibnamefont {Simakachorn}},\
  }\href@noop {} {\  (\bibinfo {year} {2023})},\ \Eprint
  {http://arxiv.org/abs/2307.02399} {arXiv:2307.02399 [astro-ph.CO]}
  \BibitemShut {NoStop}%
\bibitem [{\citenamefont {Madge}\ \emph {et~al.}(2023)\citenamefont {Madge},
  \citenamefont {Morgante}, \citenamefont {Puchades-Ib\'a\~nez}, \citenamefont
  {Ramberg}, \citenamefont {Ratzinger}, \citenamefont {Schenk},\ and\
  \citenamefont {Schwaller}}]{Madge:2023cak}%
  \BibitemOpen
  \bibfield  {author} {\bibinfo {author} {\bibfnamefont {E.}~\bibnamefont
  {Madge}}, \bibinfo {author} {\bibfnamefont {E.}~\bibnamefont {Morgante}},
  \bibinfo {author} {\bibfnamefont {C.}~\bibnamefont {Puchades-Ib\'a\~nez}},
  \bibinfo {author} {\bibfnamefont {N.}~\bibnamefont {Ramberg}}, \bibinfo
  {author} {\bibfnamefont {W.}~\bibnamefont {Ratzinger}}, \bibinfo {author}
  {\bibfnamefont {S.}~\bibnamefont {Schenk}}, \ and\ \bibinfo {author}
  {\bibfnamefont {P.}~\bibnamefont {Schwaller}},\ }\href@noop {} {\  (\bibinfo
  {year} {2023})},\ \Eprint {http://arxiv.org/abs/2306.14856} {arXiv:2306.14856
  [hep-ph]} \BibitemShut {NoStop}%
\bibitem [{\citenamefont {Kitajima}\ \emph {et~al.}(2023)\citenamefont
  {Kitajima}, \citenamefont {Lee}, \citenamefont {Murai}, \citenamefont
  {Takahashi},\ and\ \citenamefont {Yin}}]{Kitajima:2023cek}%
  \BibitemOpen
  \bibfield  {author} {\bibinfo {author} {\bibfnamefont {N.}~\bibnamefont
  {Kitajima}}, \bibinfo {author} {\bibfnamefont {J.}~\bibnamefont {Lee}},
  \bibinfo {author} {\bibfnamefont {K.}~\bibnamefont {Murai}}, \bibinfo
  {author} {\bibfnamefont {F.}~\bibnamefont {Takahashi}}, \ and\ \bibinfo
  {author} {\bibfnamefont {W.}~\bibnamefont {Yin}},\ }\href@noop {} {\
  (\bibinfo {year} {2023})},\ \Eprint {http://arxiv.org/abs/2306.17146}
  {arXiv:2306.17146 [hep-ph]} \BibitemShut {NoStop}%
\bibitem [{\citenamefont {Vagnozzi}(2023)}]{Vagnozzi:2023lwo}%
  \BibitemOpen
  \bibfield  {author} {\bibinfo {author} {\bibfnamefont {S.}~\bibnamefont
  {Vagnozzi}},\ }\href {\doibase 10.1016/j.jheap.2023.07.001} {\  (\bibinfo
  {year} {2023}),\ 10.1016/j.jheap.2023.07.001},\ \Eprint
  {http://arxiv.org/abs/2306.16912} {arXiv:2306.16912 [astro-ph.CO]}
  \BibitemShut {NoStop}%
\bibitem [{\citenamefont {Zu}\ \emph {et~al.}(2023)\citenamefont {Zu},
  \citenamefont {Zhang}, \citenamefont {Li}, \citenamefont {Gu}, \citenamefont
  {Tsai},\ and\ \citenamefont {Fan}}]{Zu:2023olm}%
  \BibitemOpen
  \bibfield  {author} {\bibinfo {author} {\bibfnamefont {L.}~\bibnamefont
  {Zu}}, \bibinfo {author} {\bibfnamefont {C.}~\bibnamefont {Zhang}}, \bibinfo
  {author} {\bibfnamefont {Y.-Y.}\ \bibnamefont {Li}}, \bibinfo {author}
  {\bibfnamefont {Y.-C.}\ \bibnamefont {Gu}}, \bibinfo {author} {\bibfnamefont
  {Y.-L.~S.}\ \bibnamefont {Tsai}}, \ and\ \bibinfo {author} {\bibfnamefont
  {Y.-Z.}\ \bibnamefont {Fan}},\ }\href@noop {} {\  (\bibinfo {year} {2023})},\
  \Eprint {http://arxiv.org/abs/2306.16769} {arXiv:2306.16769 [astro-ph.HE]}
  \BibitemShut {NoStop}%
\bibitem [{\citenamefont {Guo}\ \emph {et~al.}(2023)\citenamefont {Guo},
  \citenamefont {Khlopov}, \citenamefont {Liu}, \citenamefont {Wu},
  \citenamefont {Wu},\ and\ \citenamefont {Zhu}}]{Guo:2023hyp}%
  \BibitemOpen
  \bibfield  {author} {\bibinfo {author} {\bibfnamefont {S.-Y.}\ \bibnamefont
  {Guo}}, \bibinfo {author} {\bibfnamefont {M.}~\bibnamefont {Khlopov}},
  \bibinfo {author} {\bibfnamefont {X.}~\bibnamefont {Liu}}, \bibinfo {author}
  {\bibfnamefont {L.}~\bibnamefont {Wu}}, \bibinfo {author} {\bibfnamefont
  {Y.}~\bibnamefont {Wu}}, \ and\ \bibinfo {author} {\bibfnamefont
  {B.}~\bibnamefont {Zhu}},\ }\href@noop {} {\  (\bibinfo {year} {2023})},\
  \Eprint {http://arxiv.org/abs/2306.17022} {arXiv:2306.17022 [hep-ph]}
  \BibitemShut {NoStop}%
\bibitem [{\citenamefont {Gouttenoire}\ and\ \citenamefont
  {Vitagliano}(2023)}]{Gouttenoire:2023ftk}%
  \BibitemOpen
  \bibfield  {author} {\bibinfo {author} {\bibfnamefont {Y.}~\bibnamefont
  {Gouttenoire}}\ and\ \bibinfo {author} {\bibfnamefont {E.}~\bibnamefont
  {Vitagliano}},\ }\href@noop {} {\  (\bibinfo {year} {2023})},\ \Eprint
  {http://arxiv.org/abs/2306.17841} {arXiv:2306.17841 [gr-qc]} \BibitemShut
  {NoStop}%
\bibitem [{\citenamefont {Abe}\ and\ \citenamefont {Tada}(2023)}]{Abe:2023yrw}%
  \BibitemOpen
  \bibfield  {author} {\bibinfo {author} {\bibfnamefont {K.~T.}\ \bibnamefont
  {Abe}}\ and\ \bibinfo {author} {\bibfnamefont {Y.}~\bibnamefont {Tada}},\
  }\href@noop {} {\  (\bibinfo {year} {2023})},\ \Eprint
  {http://arxiv.org/abs/2307.01653} {arXiv:2307.01653 [astro-ph.CO]}
  \BibitemShut {NoStop}%
\bibitem [{\citenamefont {Servant}\ and\ \citenamefont
  {Simakachorn}(2023)}]{Servant:2023mwt}%
  \BibitemOpen
  \bibfield  {author} {\bibinfo {author} {\bibfnamefont {G.}~\bibnamefont
  {Servant}}\ and\ \bibinfo {author} {\bibfnamefont {P.}~\bibnamefont
  {Simakachorn}},\ }\href@noop {} {\  (\bibinfo {year} {2023})},\ \Eprint
  {http://arxiv.org/abs/2307.03121} {arXiv:2307.03121 [hep-ph]} \BibitemShut
  {NoStop}%
\bibitem [{\citenamefont {Li}(2023)}]{Li:2023tdx}%
  \BibitemOpen
  \bibfield  {author} {\bibinfo {author} {\bibfnamefont {X.-F.}\ \bibnamefont
  {Li}},\ }\href@noop {} {\  (\bibinfo {year} {2023})},\ \Eprint
  {http://arxiv.org/abs/2307.03163} {arXiv:2307.03163 [hep-ph]} \BibitemShut
  {NoStop}%
\bibitem [{\citenamefont {Geller}\ \emph {et~al.}(2023)\citenamefont {Geller},
  \citenamefont {Ghosh}, \citenamefont {Lu},\ and\ \citenamefont
  {Tsai}}]{Geller:2023shn}%
  \BibitemOpen
  \bibfield  {author} {\bibinfo {author} {\bibfnamefont {M.}~\bibnamefont
  {Geller}}, \bibinfo {author} {\bibfnamefont {S.}~\bibnamefont {Ghosh}},
  \bibinfo {author} {\bibfnamefont {S.}~\bibnamefont {Lu}}, \ and\ \bibinfo
  {author} {\bibfnamefont {Y.}~\bibnamefont {Tsai}},\ }\href@noop {} {\
  (\bibinfo {year} {2023})},\ \Eprint {http://arxiv.org/abs/2307.03724}
  {arXiv:2307.03724 [hep-ph]} \BibitemShut {NoStop}%
\bibitem [{\citenamefont {Gouttenoire}(2023)}]{Gouttenoire:2023bqy}%
  \BibitemOpen
  \bibfield  {author} {\bibinfo {author} {\bibfnamefont {Y.}~\bibnamefont
  {Gouttenoire}},\ }\href@noop {} {\  (\bibinfo {year} {2023})},\ \Eprint
  {http://arxiv.org/abs/2307.04239} {arXiv:2307.04239 [hep-ph]} \BibitemShut
  {NoStop}%
\bibitem [{\citenamefont {Salvio}(2023)}]{Salvio:2023ynn}%
  \BibitemOpen
  \bibfield  {author} {\bibinfo {author} {\bibfnamefont {A.}~\bibnamefont
  {Salvio}},\ }\href@noop {} {\  (\bibinfo {year} {2023})},\ \Eprint
  {http://arxiv.org/abs/2307.04694} {arXiv:2307.04694 [hep-ph]} \BibitemShut
  {NoStop}%
\bibitem [{\citenamefont {Babichev}\ \emph {et~al.}(2023)\citenamefont
  {Babichev}, \citenamefont {Gorbunov}, \citenamefont {Ramazanov},
  \citenamefont {Samanta},\ and\ \citenamefont {Vikman}}]{Babichev:2023pbf}%
  \BibitemOpen
  \bibfield  {author} {\bibinfo {author} {\bibfnamefont {E.}~\bibnamefont
  {Babichev}}, \bibinfo {author} {\bibfnamefont {D.}~\bibnamefont {Gorbunov}},
  \bibinfo {author} {\bibfnamefont {S.}~\bibnamefont {Ramazanov}}, \bibinfo
  {author} {\bibfnamefont {R.}~\bibnamefont {Samanta}}, \ and\ \bibinfo
  {author} {\bibfnamefont {A.}~\bibnamefont {Vikman}},\ }\href@noop {} {\
  (\bibinfo {year} {2023})},\ \Eprint {http://arxiv.org/abs/2307.04582}
  {arXiv:2307.04582 [hep-ph]} \BibitemShut {NoStop}%
\bibitem [{\citenamefont {Zhang}\ \emph {et~al.}(2023)\citenamefont {Zhang},
  \citenamefont {Cai}, \citenamefont {Su}, \citenamefont {Wang}, \citenamefont
  {Yu},\ and\ \citenamefont {Zhang}}]{Zhang:2023nrs}%
  \BibitemOpen
  \bibfield  {author} {\bibinfo {author} {\bibfnamefont {Z.}~\bibnamefont
  {Zhang}}, \bibinfo {author} {\bibfnamefont {C.}~\bibnamefont {Cai}}, \bibinfo
  {author} {\bibfnamefont {Y.-H.}\ \bibnamefont {Su}}, \bibinfo {author}
  {\bibfnamefont {S.}~\bibnamefont {Wang}}, \bibinfo {author} {\bibfnamefont
  {Z.-H.}\ \bibnamefont {Yu}}, \ and\ \bibinfo {author} {\bibfnamefont {H.-H.}\
  \bibnamefont {Zhang}},\ }\href@noop {} {\  (\bibinfo {year} {2023})},\
  \Eprint {http://arxiv.org/abs/2307.11495} {arXiv:2307.11495 [hep-ph]}
  \BibitemShut {NoStop}%
\bibitem [{\citenamefont {Ahmadvand}\ \emph {et~al.}(2023)\citenamefont
  {Ahmadvand}, \citenamefont {Bian},\ and\ \citenamefont
  {Shakeri}}]{Ahmadvand:2023lpp}%
  \BibitemOpen
  \bibfield  {author} {\bibinfo {author} {\bibfnamefont {M.}~\bibnamefont
  {Ahmadvand}}, \bibinfo {author} {\bibfnamefont {L.}~\bibnamefont {Bian}}, \
  and\ \bibinfo {author} {\bibfnamefont {S.}~\bibnamefont {Shakeri}},\
  }\href@noop {} {\  (\bibinfo {year} {2023})},\ \Eprint
  {http://arxiv.org/abs/2307.12385} {arXiv:2307.12385 [hep-ph]} \BibitemShut
  {NoStop}%
\bibitem [{\citenamefont {Franciolini}\ \emph {et~al.}(2023)\citenamefont
  {Franciolini}, \citenamefont {Iovino}, \citenamefont {Vaskonen},\ and\
  \citenamefont {Veermae}}]{Franciolini:2023pbf}%
  \BibitemOpen
  \bibfield  {author} {\bibinfo {author} {\bibfnamefont {G.}~\bibnamefont
  {Franciolini}}, \bibinfo {author} {\bibfnamefont {A.}~\bibnamefont {Iovino},
  \bibfnamefont {Junior.}}, \bibinfo {author} {\bibfnamefont {V.}~\bibnamefont
  {Vaskonen}}, \ and\ \bibinfo {author} {\bibfnamefont {H.}~\bibnamefont
  {Veermae}},\ }\href@noop {} {\  (\bibinfo {year} {2023})},\ \Eprint
  {http://arxiv.org/abs/2306.17149} {arXiv:2306.17149 [astro-ph.CO]}
  \BibitemShut {NoStop}%
\bibitem [{\citenamefont {Inomata}\ \emph {et~al.}(2023)\citenamefont
  {Inomata}, \citenamefont {Kohri},\ and\ \citenamefont
  {Terada}}]{Inomata:2023zup}%
  \BibitemOpen
  \bibfield  {author} {\bibinfo {author} {\bibfnamefont {K.}~\bibnamefont
  {Inomata}}, \bibinfo {author} {\bibfnamefont {K.}~\bibnamefont {Kohri}}, \
  and\ \bibinfo {author} {\bibfnamefont {T.}~\bibnamefont {Terada}},\
  }\href@noop {} {\  (\bibinfo {year} {2023})},\ \Eprint
  {http://arxiv.org/abs/2306.17834} {arXiv:2306.17834 [astro-ph.CO]}
  \BibitemShut {NoStop}%
\bibitem [{\citenamefont {Wang}\ \emph {et~al.}(2023)\citenamefont {Wang},
  \citenamefont {Zhao}, \citenamefont {Li},\ and\ \citenamefont
  {Zhu}}]{Wang:2023ost}%
  \BibitemOpen
  \bibfield  {author} {\bibinfo {author} {\bibfnamefont {S.}~\bibnamefont
  {Wang}}, \bibinfo {author} {\bibfnamefont {Z.-C.}\ \bibnamefont {Zhao}},
  \bibinfo {author} {\bibfnamefont {J.-P.}\ \bibnamefont {Li}}, \ and\ \bibinfo
  {author} {\bibfnamefont {Q.-H.}\ \bibnamefont {Zhu}},\ }\href@noop {} {\
  (\bibinfo {year} {2023})},\ \Eprint {http://arxiv.org/abs/2307.00572}
  {arXiv:2307.00572 [astro-ph.CO]} \BibitemShut {NoStop}%
\bibitem [{\citenamefont {Cai}\ \emph {et~al.}(2023)\citenamefont {Cai},
  \citenamefont {He}, \citenamefont {Ma}, \citenamefont {Yan},\ and\
  \citenamefont {Yuan}}]{Cai:2023dls}%
  \BibitemOpen
  \bibfield  {author} {\bibinfo {author} {\bibfnamefont {Y.-F.}\ \bibnamefont
  {Cai}}, \bibinfo {author} {\bibfnamefont {X.-C.}\ \bibnamefont {He}},
  \bibinfo {author} {\bibfnamefont {X.}~\bibnamefont {Ma}}, \bibinfo {author}
  {\bibfnamefont {S.-F.}\ \bibnamefont {Yan}}, \ and\ \bibinfo {author}
  {\bibfnamefont {G.-W.}\ \bibnamefont {Yuan}},\ }\href@noop {} {\  (\bibinfo
  {year} {2023})},\ \Eprint {http://arxiv.org/abs/2306.17822} {arXiv:2306.17822
  [gr-qc]} \BibitemShut {NoStop}%
\bibitem [{\citenamefont {Yi}\ \emph {et~al.}(2023)\citenamefont {Yi},
  \citenamefont {Gao}, \citenamefont {Gong}, \citenamefont {Wang},\ and\
  \citenamefont {Zhang}}]{Yi:2023mbm}%
  \BibitemOpen
  \bibfield  {author} {\bibinfo {author} {\bibfnamefont {Z.}~\bibnamefont
  {Yi}}, \bibinfo {author} {\bibfnamefont {Q.}~\bibnamefont {Gao}}, \bibinfo
  {author} {\bibfnamefont {Y.}~\bibnamefont {Gong}}, \bibinfo {author}
  {\bibfnamefont {Y.}~\bibnamefont {Wang}}, \ and\ \bibinfo {author}
  {\bibfnamefont {F.}~\bibnamefont {Zhang}},\ }\href@noop {} {\  (\bibinfo
  {year} {2023})},\ \Eprint {http://arxiv.org/abs/2307.02467} {arXiv:2307.02467
  [gr-qc]} \BibitemShut {NoStop}%
\bibitem [{\citenamefont {Zhu}\ \emph {et~al.}(2023)\citenamefont {Zhu},
  \citenamefont {Zhao},\ and\ \citenamefont {Wang}}]{Zhu:2023faa}%
  \BibitemOpen
  \bibfield  {author} {\bibinfo {author} {\bibfnamefont {Q.-H.}\ \bibnamefont
  {Zhu}}, \bibinfo {author} {\bibfnamefont {Z.-C.}\ \bibnamefont {Zhao}}, \
  and\ \bibinfo {author} {\bibfnamefont {S.}~\bibnamefont {Wang}},\ }\href@noop
  {} {\  (\bibinfo {year} {2023})},\ \Eprint {http://arxiv.org/abs/2307.03095}
  {arXiv:2307.03095 [astro-ph.CO]} \BibitemShut {NoStop}%
\bibitem [{\citenamefont {Firouzjahi}\ and\ \citenamefont
  {Talebian}(2023)}]{Firouzjahi:2023lzg}%
  \BibitemOpen
  \bibfield  {author} {\bibinfo {author} {\bibfnamefont {H.}~\bibnamefont
  {Firouzjahi}}\ and\ \bibinfo {author} {\bibfnamefont {A.}~\bibnamefont
  {Talebian}},\ }\href@noop {} {\  (\bibinfo {year} {2023})},\ \Eprint
  {http://arxiv.org/abs/2307.03164} {arXiv:2307.03164 [gr-qc]} \BibitemShut
  {NoStop}%
\bibitem [{\citenamefont {You}\ \emph {et~al.}(2023)\citenamefont {You},
  \citenamefont {Yi},\ and\ \citenamefont {Wu}}]{You:2023rmn}%
  \BibitemOpen
  \bibfield  {author} {\bibinfo {author} {\bibfnamefont {Z.-Q.}\ \bibnamefont
  {You}}, \bibinfo {author} {\bibfnamefont {Z.}~\bibnamefont {Yi}}, \ and\
  \bibinfo {author} {\bibfnamefont {Y.}~\bibnamefont {Wu}},\ }\href@noop {} {\
  (\bibinfo {year} {2023})},\ \Eprint {http://arxiv.org/abs/2307.04419}
  {arXiv:2307.04419 [gr-qc]} \BibitemShut {NoStop}%
\bibitem [{\citenamefont {Hosseini~Mansoori}\ \emph {et~al.}(2023)\citenamefont
  {Hosseini~Mansoori}, \citenamefont {Felegray}, \citenamefont {Talebian},\
  and\ \citenamefont {Sami}}]{HosseiniMansoori:2023mqh}%
  \BibitemOpen
  \bibfield  {author} {\bibinfo {author} {\bibfnamefont {S.~A.}\ \bibnamefont
  {Hosseini~Mansoori}}, \bibinfo {author} {\bibfnamefont {F.}~\bibnamefont
  {Felegray}}, \bibinfo {author} {\bibfnamefont {A.}~\bibnamefont {Talebian}},
  \ and\ \bibinfo {author} {\bibfnamefont {M.}~\bibnamefont {Sami}},\
  }\href@noop {} {\  (\bibinfo {year} {2023})},\ \Eprint
  {http://arxiv.org/abs/2307.06757} {arXiv:2307.06757 [astro-ph.CO]}
  \BibitemShut {NoStop}%
\bibitem [{\citenamefont {Balaji}\ \emph {et~al.}(2023)\citenamefont {Balaji},
  \citenamefont {Dom\`enech},\ and\ \citenamefont
  {Franciolini}}]{Balaji:2023ehk}%
  \BibitemOpen
  \bibfield  {author} {\bibinfo {author} {\bibfnamefont {S.}~\bibnamefont
  {Balaji}}, \bibinfo {author} {\bibfnamefont {G.}~\bibnamefont {Dom\`enech}},
  \ and\ \bibinfo {author} {\bibfnamefont {G.}~\bibnamefont {Franciolini}},\
  }\href@noop {} {\  (\bibinfo {year} {2023})},\ \Eprint
  {http://arxiv.org/abs/2307.08552} {arXiv:2307.08552 [gr-qc]} \BibitemShut
  {NoStop}%
\bibitem [{\citenamefont {Cheung}\ \emph {et~al.}(2023)\citenamefont {Cheung},
  \citenamefont {Ouseph},\ and\ \citenamefont {Tseng}}]{Cheung:2023ihl}%
  \BibitemOpen
  \bibfield  {author} {\bibinfo {author} {\bibfnamefont {K.}~\bibnamefont
  {Cheung}}, \bibinfo {author} {\bibfnamefont {C.~J.}\ \bibnamefont {Ouseph}},
  \ and\ \bibinfo {author} {\bibfnamefont {P.-Y.}\ \bibnamefont {Tseng}},\
  }\href@noop {} {\  (\bibinfo {year} {2023})},\ \Eprint
  {http://arxiv.org/abs/2307.08046} {arXiv:2307.08046 [hep-ph]} \BibitemShut
  {NoStop}%
\bibitem [{\citenamefont {Jin}\ \emph {et~al.}(2023)\citenamefont {Jin},
  \citenamefont {Chen}, \citenamefont {Yi}, \citenamefont {You}, \citenamefont
  {Liu},\ and\ \citenamefont {Wu}}]{Jin:2023wri}%
  \BibitemOpen
  \bibfield  {author} {\bibinfo {author} {\bibfnamefont {J.-H.}\ \bibnamefont
  {Jin}}, \bibinfo {author} {\bibfnamefont {Z.-C.}\ \bibnamefont {Chen}},
  \bibinfo {author} {\bibfnamefont {Z.}~\bibnamefont {Yi}}, \bibinfo {author}
  {\bibfnamefont {Z.-Q.}\ \bibnamefont {You}}, \bibinfo {author} {\bibfnamefont
  {L.}~\bibnamefont {Liu}}, \ and\ \bibinfo {author} {\bibfnamefont
  {Y.}~\bibnamefont {Wu}},\ }\href@noop {} {\  (\bibinfo {year} {2023})},\
  \Eprint {http://arxiv.org/abs/2307.08687} {arXiv:2307.08687 [astro-ph.CO]}
  \BibitemShut {NoStop}%
\bibitem [{\citenamefont {Ananda}\ \emph {et~al.}(2007)\citenamefont {Ananda},
  \citenamefont {Clarkson},\ and\ \citenamefont {Wands}}]{Ananda:2006af}%
  \BibitemOpen
  \bibfield  {author} {\bibinfo {author} {\bibfnamefont {K.~N.}\ \bibnamefont
  {Ananda}}, \bibinfo {author} {\bibfnamefont {C.}~\bibnamefont {Clarkson}}, \
  and\ \bibinfo {author} {\bibfnamefont {D.}~\bibnamefont {Wands}},\ }\href
  {\doibase 10.1103/PhysRevD.75.123518} {\bibfield  {journal} {\bibinfo
  {journal} {Phys. Rev. D}\ }\textbf {\bibinfo {volume} {75}},\ \bibinfo
  {pages} {123518} (\bibinfo {year} {2007})},\ \Eprint
  {http://arxiv.org/abs/gr-qc/0612013} {arXiv:gr-qc/0612013} \BibitemShut
  {NoStop}%
\bibitem [{\citenamefont {Baumann}\ \emph {et~al.}(2007)\citenamefont
  {Baumann}, \citenamefont {Steinhardt}, \citenamefont {Takahashi},\ and\
  \citenamefont {Ichiki}}]{Baumann:2007zm}%
  \BibitemOpen
  \bibfield  {author} {\bibinfo {author} {\bibfnamefont {D.}~\bibnamefont
  {Baumann}}, \bibinfo {author} {\bibfnamefont {P.~J.}\ \bibnamefont
  {Steinhardt}}, \bibinfo {author} {\bibfnamefont {K.}~\bibnamefont
  {Takahashi}}, \ and\ \bibinfo {author} {\bibfnamefont {K.}~\bibnamefont
  {Ichiki}},\ }\href {\doibase 10.1103/PhysRevD.76.084019} {\bibfield
  {journal} {\bibinfo  {journal} {Phys. Rev. D}\ }\textbf {\bibinfo {volume}
  {76}},\ \bibinfo {pages} {084019} (\bibinfo {year} {2007})},\ \Eprint
  {http://arxiv.org/abs/hep-th/0703290} {arXiv:hep-th/0703290} \BibitemShut
  {NoStop}%
\bibitem [{\citenamefont {Alabidi}\ \emph {et~al.}(2012)\citenamefont
  {Alabidi}, \citenamefont {Kohri}, \citenamefont {Sasaki},\ and\ \citenamefont
  {Sendouda}}]{Alabidi:2012ex}%
  \BibitemOpen
  \bibfield  {author} {\bibinfo {author} {\bibfnamefont {L.}~\bibnamefont
  {Alabidi}}, \bibinfo {author} {\bibfnamefont {K.}~\bibnamefont {Kohri}},
  \bibinfo {author} {\bibfnamefont {M.}~\bibnamefont {Sasaki}}, \ and\ \bibinfo
  {author} {\bibfnamefont {Y.}~\bibnamefont {Sendouda}},\ }\href {\doibase
  10.1088/1475-7516/2012/09/017} {\bibfield  {journal} {\bibinfo  {journal}
  {JCAP}\ }\textbf {\bibinfo {volume} {09}},\ \bibinfo {pages} {017} (\bibinfo
  {year} {2012})},\ \Eprint {http://arxiv.org/abs/1203.4663} {arXiv:1203.4663
  [astro-ph.CO]} \BibitemShut {NoStop}%
\bibitem [{\citenamefont {Espinosa}\ \emph {et~al.}(2018)\citenamefont
  {Espinosa}, \citenamefont {Racco},\ and\ \citenamefont
  {Riotto}}]{Espinosa:2018eve}%
  \BibitemOpen
  \bibfield  {author} {\bibinfo {author} {\bibfnamefont {J.~R.}\ \bibnamefont
  {Espinosa}}, \bibinfo {author} {\bibfnamefont {D.}~\bibnamefont {Racco}}, \
  and\ \bibinfo {author} {\bibfnamefont {A.}~\bibnamefont {Riotto}},\ }\href
  {\doibase 10.1088/1475-7516/2018/09/012} {\bibfield  {journal} {\bibinfo
  {journal} {JCAP}\ }\textbf {\bibinfo {volume} {09}},\ \bibinfo {pages} {012}
  (\bibinfo {year} {2018})},\ \Eprint {http://arxiv.org/abs/1804.07732}
  {arXiv:1804.07732 [hep-ph]} \BibitemShut {NoStop}%
\bibitem [{\citenamefont {Kohri}\ and\ \citenamefont
  {Terada}(2018)}]{Kohri:2018awv}%
  \BibitemOpen
  \bibfield  {author} {\bibinfo {author} {\bibfnamefont {K.}~\bibnamefont
  {Kohri}}\ and\ \bibinfo {author} {\bibfnamefont {T.}~\bibnamefont {Terada}},\
  }\href {\doibase 10.1103/PhysRevD.97.123532} {\bibfield  {journal} {\bibinfo
  {journal} {Phys. Rev. D}\ }\textbf {\bibinfo {volume} {97}},\ \bibinfo
  {pages} {123532} (\bibinfo {year} {2018})},\ \Eprint
  {http://arxiv.org/abs/1804.08577} {arXiv:1804.08577 [gr-qc]} \BibitemShut
  {NoStop}%
\bibitem [{\citenamefont {Caprini}\ and\ \citenamefont
  {Figueroa}(2018)}]{Caprini:2018mtu}%
  \BibitemOpen
  \bibfield  {author} {\bibinfo {author} {\bibfnamefont {C.}~\bibnamefont
  {Caprini}}\ and\ \bibinfo {author} {\bibfnamefont {D.~G.}\ \bibnamefont
  {Figueroa}},\ }\href {\doibase 10.1088/1361-6382/aac608} {\bibfield
  {journal} {\bibinfo  {journal} {Class. Quant. Grav.}\ }\textbf {\bibinfo
  {volume} {35}},\ \bibinfo {pages} {163001} (\bibinfo {year} {2018})},\
  \Eprint {http://arxiv.org/abs/1801.04268} {arXiv:1801.04268 [astro-ph.CO]}
  \BibitemShut {NoStop}%
\bibitem [{\citenamefont {Akrami}\ \emph {et~al.}(2020)\citenamefont {Akrami}
  \emph {et~al.}}]{Planck:2018jri}%
  \BibitemOpen
  \bibfield  {author} {\bibinfo {author} {\bibfnamefont {Y.}~\bibnamefont
  {Akrami}} \emph {et~al.} (\bibinfo {collaboration} {Planck}),\ }\href
  {\doibase 10.1051/0004-6361/201833887} {\bibfield  {journal} {\bibinfo
  {journal} {Astron. Astrophys.}\ }\textbf {\bibinfo {volume} {641}},\ \bibinfo
  {pages} {A10} (\bibinfo {year} {2020})},\ \Eprint
  {http://arxiv.org/abs/1807.06211} {arXiv:1807.06211 [astro-ph.CO]}
  \BibitemShut {NoStop}%
\bibitem [{\citenamefont {Tristram}\ \emph {et~al.}(2022)\citenamefont
  {Tristram} \emph {et~al.}}]{Tristram:2021tvh}%
  \BibitemOpen
  \bibfield  {author} {\bibinfo {author} {\bibfnamefont {M.}~\bibnamefont
  {Tristram}} \emph {et~al.},\ }\href {\doibase 10.1103/PhysRevD.105.083524}
  {\bibfield  {journal} {\bibinfo  {journal} {Phys. Rev. D}\ }\textbf {\bibinfo
  {volume} {105}},\ \bibinfo {pages} {083524} (\bibinfo {year} {2022})},\
  \Eprint {http://arxiv.org/abs/2112.07961} {arXiv:2112.07961 [astro-ph.CO]}
  \BibitemShut {NoStop}%
\bibitem [{\citenamefont {Saito}\ and\ \citenamefont
  {Yokoyama}(2009)}]{Saito:2008jc}%
  \BibitemOpen
  \bibfield  {author} {\bibinfo {author} {\bibfnamefont {R.}~\bibnamefont
  {Saito}}\ and\ \bibinfo {author} {\bibfnamefont {J.}~\bibnamefont
  {Yokoyama}},\ }\href {\doibase 10.1103/PhysRevLett.102.161101} {\bibfield
  {journal} {\bibinfo  {journal} {Phys. Rev. Lett.}\ }\textbf {\bibinfo
  {volume} {102}},\ \bibinfo {pages} {161101} (\bibinfo {year} {2009})},\
  \bibinfo {note} {[Erratum: Phys.Rev.Lett. 107, 069901 (2011)]},\ \Eprint
  {http://arxiv.org/abs/0812.4339} {arXiv:0812.4339 [astro-ph]} \BibitemShut
  {NoStop}%
\bibitem [{\citenamefont {Pi}\ and\ \citenamefont {Sasaki}(2020)}]{Pi:2020otn}%
  \BibitemOpen
  \bibfield  {author} {\bibinfo {author} {\bibfnamefont {S.}~\bibnamefont
  {Pi}}\ and\ \bibinfo {author} {\bibfnamefont {M.}~\bibnamefont {Sasaki}},\
  }\href {\doibase 10.1088/1475-7516/2020/09/037} {\bibfield  {journal}
  {\bibinfo  {journal} {JCAP}\ }\textbf {\bibinfo {volume} {09}},\ \bibinfo
  {pages} {037} (\bibinfo {year} {2020})},\ \Eprint
  {http://arxiv.org/abs/2005.12306} {arXiv:2005.12306 [gr-qc]} \BibitemShut
  {NoStop}%
\bibitem [{\citenamefont {Dom\`enech}(2021)}]{Domenech:2021ztg}%
  \BibitemOpen
  \bibfield  {author} {\bibinfo {author} {\bibfnamefont {G.}~\bibnamefont
  {Dom\`enech}},\ }\href {\doibase 10.3390/universe7110398} {\bibfield
  {journal} {\bibinfo  {journal} {Universe}\ }\textbf {\bibinfo {volume} {7}},\
  \bibinfo {pages} {398} (\bibinfo {year} {2021})},\ \Eprint
  {http://arxiv.org/abs/2109.01398} {arXiv:2109.01398 [gr-qc]} \BibitemShut
  {NoStop}%
\bibitem [{\citenamefont {Sasaki}\ \emph {et~al.}(2018)\citenamefont {Sasaki},
  \citenamefont {Suyama}, \citenamefont {Tanaka},\ and\ \citenamefont
  {Yokoyama}}]{Sasaki:2018dmp}%
  \BibitemOpen
  \bibfield  {author} {\bibinfo {author} {\bibfnamefont {M.}~\bibnamefont
  {Sasaki}}, \bibinfo {author} {\bibfnamefont {T.}~\bibnamefont {Suyama}},
  \bibinfo {author} {\bibfnamefont {T.}~\bibnamefont {Tanaka}}, \ and\ \bibinfo
  {author} {\bibfnamefont {S.}~\bibnamefont {Yokoyama}},\ }\href {\doibase
  10.1088/1361-6382/aaa7b4} {\bibfield  {journal} {\bibinfo  {journal} {Class.
  Quant. Grav.}\ }\textbf {\bibinfo {volume} {35}},\ \bibinfo {pages} {063001}
  (\bibinfo {year} {2018})},\ \Eprint {http://arxiv.org/abs/1801.05235}
  {arXiv:1801.05235 [astro-ph.CO]} \BibitemShut {NoStop}%
\bibitem [{\citenamefont {Yoo}\ \emph {et~al.}(2018)\citenamefont {Yoo},
  \citenamefont {Harada}, \citenamefont {Garriga},\ and\ \citenamefont
  {Kohri}}]{Yoo:2018kvb}%
  \BibitemOpen
  \bibfield  {author} {\bibinfo {author} {\bibfnamefont {C.-M.}\ \bibnamefont
  {Yoo}}, \bibinfo {author} {\bibfnamefont {T.}~\bibnamefont {Harada}},
  \bibinfo {author} {\bibfnamefont {J.}~\bibnamefont {Garriga}}, \ and\
  \bibinfo {author} {\bibfnamefont {K.}~\bibnamefont {Kohri}},\ }\href
  {\doibase 10.1093/ptep/pty120} {\bibfield  {journal} {\bibinfo  {journal}
  {PTEP}\ }\textbf {\bibinfo {volume} {2018}},\ \bibinfo {pages} {123E01}
  (\bibinfo {year} {2018})},\ \Eprint {http://arxiv.org/abs/1805.03946}
  {arXiv:1805.03946 [astro-ph.CO]} \BibitemShut {NoStop}%
\bibitem [{\citenamefont {Germani}\ and\ \citenamefont
  {Musco}(2019)}]{Germani:2018jgr}%
  \BibitemOpen
  \bibfield  {author} {\bibinfo {author} {\bibfnamefont {C.}~\bibnamefont
  {Germani}}\ and\ \bibinfo {author} {\bibfnamefont {I.}~\bibnamefont
  {Musco}},\ }\href {\doibase 10.1103/PhysRevLett.122.141302} {\bibfield
  {journal} {\bibinfo  {journal} {Phys. Rev. Lett.}\ }\textbf {\bibinfo
  {volume} {122}},\ \bibinfo {pages} {141302} (\bibinfo {year} {2019})},\
  \Eprint {http://arxiv.org/abs/1805.04087} {arXiv:1805.04087 [astro-ph.CO]}
  \BibitemShut {NoStop}%
\bibitem [{\citenamefont {Germani}\ and\ \citenamefont
  {Sheth}(2020)}]{Germani:2019zez}%
  \BibitemOpen
  \bibfield  {author} {\bibinfo {author} {\bibfnamefont {C.}~\bibnamefont
  {Germani}}\ and\ \bibinfo {author} {\bibfnamefont {R.~K.}\ \bibnamefont
  {Sheth}},\ }\href {\doibase 10.1103/PhysRevD.101.063520} {\bibfield
  {journal} {\bibinfo  {journal} {Phys. Rev. D}\ }\textbf {\bibinfo {volume}
  {101}},\ \bibinfo {pages} {063520} (\bibinfo {year} {2020})},\ \Eprint
  {http://arxiv.org/abs/1912.07072} {arXiv:1912.07072 [astro-ph.CO]}
  \BibitemShut {NoStop}%
\bibitem [{\citenamefont {Saito}\ \emph {et~al.}(2008)\citenamefont {Saito},
  \citenamefont {Yokoyama},\ and\ \citenamefont {Nagata}}]{Saito:2008em}%
  \BibitemOpen
  \bibfield  {author} {\bibinfo {author} {\bibfnamefont {R.}~\bibnamefont
  {Saito}}, \bibinfo {author} {\bibfnamefont {J.}~\bibnamefont {Yokoyama}}, \
  and\ \bibinfo {author} {\bibfnamefont {R.}~\bibnamefont {Nagata}},\ }\href
  {\doibase 10.1088/1475-7516/2008/06/024} {\bibfield  {journal} {\bibinfo
  {journal} {JCAP}\ }\textbf {\bibinfo {volume} {06}},\ \bibinfo {pages} {024}
  (\bibinfo {year} {2008})},\ \Eprint {http://arxiv.org/abs/0804.3470}
  {arXiv:0804.3470 [astro-ph]} \BibitemShut {NoStop}%
\bibitem [{\citenamefont {Young}\ and\ \citenamefont
  {Byrnes}(2013)}]{Young:2013oia}%
  \BibitemOpen
  \bibfield  {author} {\bibinfo {author} {\bibfnamefont {S.}~\bibnamefont
  {Young}}\ and\ \bibinfo {author} {\bibfnamefont {C.~T.}\ \bibnamefont
  {Byrnes}},\ }\href {\doibase 10.1088/1475-7516/2013/08/052} {\bibfield
  {journal} {\bibinfo  {journal} {JCAP}\ }\textbf {\bibinfo {volume} {08}},\
  \bibinfo {pages} {052} (\bibinfo {year} {2013})},\ \Eprint
  {http://arxiv.org/abs/1307.4995} {arXiv:1307.4995 [astro-ph.CO]} \BibitemShut
  {NoStop}%
\bibitem [{\citenamefont {Franciolini}\ \emph {et~al.}(2018)\citenamefont
  {Franciolini}, \citenamefont {Kehagias}, \citenamefont {Matarrese},\ and\
  \citenamefont {Riotto}}]{Franciolini:2018vbk}%
  \BibitemOpen
  \bibfield  {author} {\bibinfo {author} {\bibfnamefont {G.}~\bibnamefont
  {Franciolini}}, \bibinfo {author} {\bibfnamefont {A.}~\bibnamefont
  {Kehagias}}, \bibinfo {author} {\bibfnamefont {S.}~\bibnamefont {Matarrese}},
  \ and\ \bibinfo {author} {\bibfnamefont {A.}~\bibnamefont {Riotto}},\ }\href
  {\doibase 10.1088/1475-7516/2018/03/016} {\bibfield  {journal} {\bibinfo
  {journal} {JCAP}\ }\textbf {\bibinfo {volume} {03}},\ \bibinfo {pages} {016}
  (\bibinfo {year} {2018})},\ \Eprint {http://arxiv.org/abs/1801.09415}
  {arXiv:1801.09415 [astro-ph.CO]} \BibitemShut {NoStop}%
\bibitem [{\citenamefont {Atal}\ and\ \citenamefont
  {Germani}(2019)}]{Atal:2018neu}%
  \BibitemOpen
  \bibfield  {author} {\bibinfo {author} {\bibfnamefont {V.}~\bibnamefont
  {Atal}}\ and\ \bibinfo {author} {\bibfnamefont {C.}~\bibnamefont {Germani}},\
  }\href {\doibase 10.1016/j.dark.2019.100275} {\bibfield  {journal} {\bibinfo
  {journal} {Phys. Dark Univ.}\ }\textbf {\bibinfo {volume} {24}},\ \bibinfo
  {pages} {100275} (\bibinfo {year} {2019})},\ \Eprint
  {http://arxiv.org/abs/1811.07857} {arXiv:1811.07857 [astro-ph.CO]}
  \BibitemShut {NoStop}%
\bibitem [{\citenamefont {Cai}\ \emph {et~al.}(2019)\citenamefont {Cai},
  \citenamefont {Pi},\ and\ \citenamefont {Sasaki}}]{Cai:2018dig}%
  \BibitemOpen
  \bibfield  {author} {\bibinfo {author} {\bibfnamefont {R.-g.}\ \bibnamefont
  {Cai}}, \bibinfo {author} {\bibfnamefont {S.}~\bibnamefont {Pi}}, \ and\
  \bibinfo {author} {\bibfnamefont {M.}~\bibnamefont {Sasaki}},\ }\href
  {\doibase 10.1103/PhysRevLett.122.201101} {\bibfield  {journal} {\bibinfo
  {journal} {Phys. Rev. Lett.}\ }\textbf {\bibinfo {volume} {122}},\ \bibinfo
  {pages} {201101} (\bibinfo {year} {2019})},\ \Eprint
  {http://arxiv.org/abs/1810.11000} {arXiv:1810.11000 [astro-ph.CO]}
  \BibitemShut {NoStop}%
\bibitem [{\citenamefont {Passaglia}\ \emph {et~al.}(2019)\citenamefont
  {Passaglia}, \citenamefont {Hu},\ and\ \citenamefont
  {Motohashi}}]{Passaglia:2018ixg}%
  \BibitemOpen
  \bibfield  {author} {\bibinfo {author} {\bibfnamefont {S.}~\bibnamefont
  {Passaglia}}, \bibinfo {author} {\bibfnamefont {W.}~\bibnamefont {Hu}}, \
  and\ \bibinfo {author} {\bibfnamefont {H.}~\bibnamefont {Motohashi}},\ }\href
  {\doibase 10.1103/PhysRevD.99.043536} {\bibfield  {journal} {\bibinfo
  {journal} {Phys. Rev. D}\ }\textbf {\bibinfo {volume} {99}},\ \bibinfo
  {pages} {043536} (\bibinfo {year} {2019})},\ \Eprint
  {http://arxiv.org/abs/1812.08243} {arXiv:1812.08243 [astro-ph.CO]}
  \BibitemShut {NoStop}%
\bibitem [{\citenamefont {De~Luca}\ \emph {et~al.}(2019)\citenamefont
  {De~Luca}, \citenamefont {Franciolini}, \citenamefont {Kehagias},
  \citenamefont {Peloso}, \citenamefont {Riotto},\ and\ \citenamefont
  {\"Unal}}]{DeLuca:2019qsy}%
  \BibitemOpen
  \bibfield  {author} {\bibinfo {author} {\bibfnamefont {V.}~\bibnamefont
  {De~Luca}}, \bibinfo {author} {\bibfnamefont {G.}~\bibnamefont
  {Franciolini}}, \bibinfo {author} {\bibfnamefont {A.}~\bibnamefont
  {Kehagias}}, \bibinfo {author} {\bibfnamefont {M.}~\bibnamefont {Peloso}},
  \bibinfo {author} {\bibfnamefont {A.}~\bibnamefont {Riotto}}, \ and\ \bibinfo
  {author} {\bibfnamefont {C.}~\bibnamefont {\"Unal}},\ }\href {\doibase
  10.1088/1475-7516/2019/07/048} {\bibfield  {journal} {\bibinfo  {journal}
  {JCAP}\ }\textbf {\bibinfo {volume} {07}},\ \bibinfo {pages} {048} (\bibinfo
  {year} {2019})},\ \Eprint {http://arxiv.org/abs/1904.00970} {arXiv:1904.00970
  [astro-ph.CO]} \BibitemShut {NoStop}%
\bibitem [{\citenamefont {Atal}\ \emph {et~al.}(2019)\citenamefont {Atal},
  \citenamefont {Garriga},\ and\ \citenamefont
  {Marcos-Caballero}}]{Atal:2019cdz}%
  \BibitemOpen
  \bibfield  {author} {\bibinfo {author} {\bibfnamefont {V.}~\bibnamefont
  {Atal}}, \bibinfo {author} {\bibfnamefont {J.}~\bibnamefont {Garriga}}, \
  and\ \bibinfo {author} {\bibfnamefont {A.}~\bibnamefont {Marcos-Caballero}},\
  }\href {\doibase 10.1088/1475-7516/2019/09/073} {\bibfield  {journal}
  {\bibinfo  {journal} {JCAP}\ }\textbf {\bibinfo {volume} {09}},\ \bibinfo
  {pages} {073} (\bibinfo {year} {2019})},\ \Eprint
  {http://arxiv.org/abs/1905.13202} {arXiv:1905.13202 [astro-ph.CO]}
  \BibitemShut {NoStop}%
\bibitem [{\citenamefont {Suyama}\ and\ \citenamefont
  {Yokoyama}(2019)}]{Suyama:2019cst}%
  \BibitemOpen
  \bibfield  {author} {\bibinfo {author} {\bibfnamefont {T.}~\bibnamefont
  {Suyama}}\ and\ \bibinfo {author} {\bibfnamefont {S.}~\bibnamefont
  {Yokoyama}},\ }\href {\doibase 10.1093/ptep/ptz105} {\bibfield  {journal}
  {\bibinfo  {journal} {PTEP}\ }\textbf {\bibinfo {volume} {2019}},\ \bibinfo
  {pages} {103E02} (\bibinfo {year} {2019})},\ \Eprint
  {http://arxiv.org/abs/1906.04958} {arXiv:1906.04958 [astro-ph.CO]}
  \BibitemShut {NoStop}%
\bibitem [{\citenamefont {Yoo}\ \emph {et~al.}(2019)\citenamefont {Yoo},
  \citenamefont {Gong},\ and\ \citenamefont {Yokoyama}}]{Yoo:2019pma}%
  \BibitemOpen
  \bibfield  {author} {\bibinfo {author} {\bibfnamefont {C.-M.}\ \bibnamefont
  {Yoo}}, \bibinfo {author} {\bibfnamefont {J.-O.}\ \bibnamefont {Gong}}, \
  and\ \bibinfo {author} {\bibfnamefont {S.}~\bibnamefont {Yokoyama}},\ }\href
  {\doibase 10.1088/1475-7516/2019/09/033} {\bibfield  {journal} {\bibinfo
  {journal} {JCAP}\ }\textbf {\bibinfo {volume} {09}},\ \bibinfo {pages} {033}
  (\bibinfo {year} {2019})},\ \Eprint {http://arxiv.org/abs/1906.06790}
  {arXiv:1906.06790 [astro-ph.CO]} \BibitemShut {NoStop}%
\bibitem [{\citenamefont {Atal}\ \emph {et~al.}(2020)\citenamefont {Atal},
  \citenamefont {Cid}, \citenamefont {Escriv\`a},\ and\ \citenamefont
  {Garriga}}]{Atal:2019erb}%
  \BibitemOpen
  \bibfield  {author} {\bibinfo {author} {\bibfnamefont {V.}~\bibnamefont
  {Atal}}, \bibinfo {author} {\bibfnamefont {J.}~\bibnamefont {Cid}}, \bibinfo
  {author} {\bibfnamefont {A.}~\bibnamefont {Escriv\`a}}, \ and\ \bibinfo
  {author} {\bibfnamefont {J.}~\bibnamefont {Garriga}},\ }\href {\doibase
  10.1088/1475-7516/2020/05/022} {\bibfield  {journal} {\bibinfo  {journal}
  {JCAP}\ }\textbf {\bibinfo {volume} {05}},\ \bibinfo {pages} {022} (\bibinfo
  {year} {2020})},\ \Eprint {http://arxiv.org/abs/1908.11357} {arXiv:1908.11357
  [astro-ph.CO]} \BibitemShut {NoStop}%
\bibitem [{\citenamefont {Ezquiaga}\ \emph {et~al.}(2020)\citenamefont
  {Ezquiaga}, \citenamefont {Garc\'\i{}a-Bellido},\ and\ \citenamefont
  {Vennin}}]{Ezquiaga:2019ftu}%
  \BibitemOpen
  \bibfield  {author} {\bibinfo {author} {\bibfnamefont {J.~M.}\ \bibnamefont
  {Ezquiaga}}, \bibinfo {author} {\bibfnamefont {J.}~\bibnamefont
  {Garc\'\i{}a-Bellido}}, \ and\ \bibinfo {author} {\bibfnamefont
  {V.}~\bibnamefont {Vennin}},\ }\href {\doibase 10.1088/1475-7516/2020/03/029}
  {\bibfield  {journal} {\bibinfo  {journal} {JCAP}\ }\textbf {\bibinfo
  {volume} {03}},\ \bibinfo {pages} {029} (\bibinfo {year} {2020})},\ \Eprint
  {http://arxiv.org/abs/1912.05399} {arXiv:1912.05399 [astro-ph.CO]}
  \BibitemShut {NoStop}%
\bibitem [{\citenamefont {Cai}\ \emph {et~al.}(2022)\citenamefont {Cai},
  \citenamefont {Ma}, \citenamefont {Sasaki}, \citenamefont {Wang},\ and\
  \citenamefont {Zhou}}]{Cai:2021zsp}%
  \BibitemOpen
  \bibfield  {author} {\bibinfo {author} {\bibfnamefont {Y.-F.}\ \bibnamefont
  {Cai}}, \bibinfo {author} {\bibfnamefont {X.-H.}\ \bibnamefont {Ma}},
  \bibinfo {author} {\bibfnamefont {M.}~\bibnamefont {Sasaki}}, \bibinfo
  {author} {\bibfnamefont {D.-G.}\ \bibnamefont {Wang}}, \ and\ \bibinfo
  {author} {\bibfnamefont {Z.}~\bibnamefont {Zhou}},\ }\href {\doibase
  10.1016/j.physletb.2022.137461} {\bibfield  {journal} {\bibinfo  {journal}
  {Phys. Lett. B}\ }\textbf {\bibinfo {volume} {834}},\ \bibinfo {pages}
  {137461} (\bibinfo {year} {2022})},\ \Eprint
  {http://arxiv.org/abs/2112.13836} {arXiv:2112.13836 [astro-ph.CO]}
  \BibitemShut {NoStop}%
\bibitem [{\citenamefont {Pi}\ and\ \citenamefont {Sasaki}(2023)}]{Pi:2022ysn}%
  \BibitemOpen
  \bibfield  {author} {\bibinfo {author} {\bibfnamefont {S.}~\bibnamefont
  {Pi}}\ and\ \bibinfo {author} {\bibfnamefont {M.}~\bibnamefont {Sasaki}},\
  }\href {\doibase 10.1103/PhysRevLett.131.011002} {\bibfield  {journal}
  {\bibinfo  {journal} {Phys. Rev. Lett.}\ }\textbf {\bibinfo {volume} {131}},\
  \bibinfo {pages} {011002} (\bibinfo {year} {2023})},\ \Eprint
  {http://arxiv.org/abs/2211.13932} {arXiv:2211.13932 [astro-ph.CO]}
  \BibitemShut {NoStop}%
\bibitem [{\citenamefont {Liu}\ \emph {et~al.}(2023)\citenamefont {Liu},
  \citenamefont {Chen},\ and\ \citenamefont {Huang}}]{Liu:2023ymk}%
  \BibitemOpen
  \bibfield  {author} {\bibinfo {author} {\bibfnamefont {L.}~\bibnamefont
  {Liu}}, \bibinfo {author} {\bibfnamefont {Z.-C.}\ \bibnamefont {Chen}}, \
  and\ \bibinfo {author} {\bibfnamefont {Q.-G.}\ \bibnamefont {Huang}},\
  }\href@noop {} {\  (\bibinfo {year} {2023})},\ \Eprint
  {http://arxiv.org/abs/2307.01102} {arXiv:2307.01102 [astro-ph.CO]}
  \BibitemShut {NoStop}%
\bibitem [{\citenamefont {Gorji}\ and\ \citenamefont
  {Sasaki}(2023)}]{Gorji:2023ziy}%
  \BibitemOpen
  \bibfield  {author} {\bibinfo {author} {\bibfnamefont {M.~A.}\ \bibnamefont
  {Gorji}}\ and\ \bibinfo {author} {\bibfnamefont {M.}~\bibnamefont {Sasaki}},\
  }\href@noop {} {\  (\bibinfo {year} {2023})},\ \Eprint
  {http://arxiv.org/abs/2302.14080} {arXiv:2302.14080 [gr-qc]} \BibitemShut
  {NoStop}%
\bibitem [{\citenamefont {Bordin}\ \emph {et~al.}(2018)\citenamefont {Bordin},
  \citenamefont {Creminelli}, \citenamefont {Khmelnitsky},\ and\ \citenamefont
  {Senatore}}]{Bordin:2018pca}%
  \BibitemOpen
  \bibfield  {author} {\bibinfo {author} {\bibfnamefont {L.}~\bibnamefont
  {Bordin}}, \bibinfo {author} {\bibfnamefont {P.}~\bibnamefont {Creminelli}},
  \bibinfo {author} {\bibfnamefont {A.}~\bibnamefont {Khmelnitsky}}, \ and\
  \bibinfo {author} {\bibfnamefont {L.}~\bibnamefont {Senatore}},\ }\href
  {\doibase 10.1088/1475-7516/2018/10/013} {\bibfield  {journal} {\bibinfo
  {journal} {JCAP}\ }\textbf {\bibinfo {volume} {10}},\ \bibinfo {pages} {013}
  (\bibinfo {year} {2018})},\ \Eprint {http://arxiv.org/abs/1806.10587}
  {arXiv:1806.10587 [hep-th]} \BibitemShut {NoStop}%
\bibitem [{\citenamefont {Maleknejad}\ and\ \citenamefont
  {Sheikh-Jabbari}(2013)}]{Maleknejad:2011jw}%
  \BibitemOpen
  \bibfield  {author} {\bibinfo {author} {\bibfnamefont {A.}~\bibnamefont
  {Maleknejad}}\ and\ \bibinfo {author} {\bibfnamefont {M.~M.}\ \bibnamefont
  {Sheikh-Jabbari}},\ }\href {\doibase 10.1016/j.physletb.2013.05.001}
  {\bibfield  {journal} {\bibinfo  {journal} {Phys. Lett. B}\ }\textbf
  {\bibinfo {volume} {723}},\ \bibinfo {pages} {224} (\bibinfo {year}
  {2013})},\ \Eprint {http://arxiv.org/abs/1102.1513} {arXiv:1102.1513
  [hep-ph]} \BibitemShut {NoStop}%
\bibitem [{\citenamefont {Dimastrogiovanni}\ \emph {et~al.}(2017)\citenamefont
  {Dimastrogiovanni}, \citenamefont {Fasiello},\ and\ \citenamefont
  {Fujita}}]{Dimastrogiovanni:2016fuu}%
  \BibitemOpen
  \bibfield  {author} {\bibinfo {author} {\bibfnamefont {E.}~\bibnamefont
  {Dimastrogiovanni}}, \bibinfo {author} {\bibfnamefont {M.}~\bibnamefont
  {Fasiello}}, \ and\ \bibinfo {author} {\bibfnamefont {T.}~\bibnamefont
  {Fujita}},\ }\href {\doibase 10.1088/1475-7516/2017/01/019} {\bibfield
  {journal} {\bibinfo  {journal} {JCAP}\ }\textbf {\bibinfo {volume} {01}},\
  \bibinfo {pages} {019} (\bibinfo {year} {2017})},\ \Eprint
  {http://arxiv.org/abs/1608.04216} {arXiv:1608.04216 [astro-ph.CO]}
  \BibitemShut {NoStop}%
\bibitem [{\citenamefont {Aghanim}\ \emph {et~al.}(2020)\citenamefont {Aghanim}
  \emph {et~al.}}]{Planck:2018vyg}%
  \BibitemOpen
  \bibfield  {author} {\bibinfo {author} {\bibfnamefont {N.}~\bibnamefont
  {Aghanim}} \emph {et~al.} (\bibinfo {collaboration} {Planck}),\ }\href
  {\doibase 10.1051/0004-6361/201833910} {\bibfield  {journal} {\bibinfo
  {journal} {Astron. Astrophys.}\ }\textbf {\bibinfo {volume} {641}},\ \bibinfo
  {pages} {A6} (\bibinfo {year} {2020})},\ \bibinfo {note} {[Erratum:
  Astron.Astrophys. 652, C4 (2021)]},\ \Eprint
  {http://arxiv.org/abs/1807.06209} {arXiv:1807.06209 [astro-ph.CO]}
  \BibitemShut {NoStop}%
\bibitem [{\citenamefont {Harada}\ \emph {et~al.}(2013)\citenamefont {Harada},
  \citenamefont {Yoo},\ and\ \citenamefont {Kohri}}]{Harada:2013epa}%
  \BibitemOpen
  \bibfield  {author} {\bibinfo {author} {\bibfnamefont {T.}~\bibnamefont
  {Harada}}, \bibinfo {author} {\bibfnamefont {C.-M.}\ \bibnamefont {Yoo}}, \
  and\ \bibinfo {author} {\bibfnamefont {K.}~\bibnamefont {Kohri}},\ }\href
  {\doibase 10.1103/PhysRevD.88.084051} {\bibfield  {journal} {\bibinfo
  {journal} {Phys. Rev. D}\ }\textbf {\bibinfo {volume} {88}},\ \bibinfo
  {pages} {084051} (\bibinfo {year} {2013})},\ \bibinfo {note} {[Erratum:
  Phys.Rev.D 89, 029903 (2014)]},\ \Eprint {http://arxiv.org/abs/1309.4201}
  {arXiv:1309.4201 [astro-ph.CO]} \BibitemShut {NoStop}%
\end{thebibliography}%

\end{document}